\documentclass{vgtc}                          

\graphicspath{{figures/}{pictures/}{images/}{./}} 

\usepackage{times}                     
\usepackage{multirow}

\usepackage{tabu}                      
\usepackage{booktabs}                  
\usepackage{lipsum}                    
\usepackage{mwe}                       

\usepackage{mathptmx}                  

\onlineid{3689}

\vgtccategory{Research}

\vgtcinsertpkg

\title{Multimodal 3D Fusion and In-Situ Learning for Spatially Aware AI}

\author{Chengyuan Xu\thanks{e-mail: cxu@ucsb.edu. Equal contribution.} %
\and Radha Kumaran\thanks{e-mail: rkumaran@ucsb.edu. Equal contribution.} %
\and Noah Stier\thanks{e-mail: noahstier@ucsb.edu} %
\and Kangyou Yu\thanks{e-mail: kangyouyu@ucsb.edu} %
\and Tobias Höllerer\thanks{e-mail: holl@cs.ucsb.edu}} %
\affiliation{\scriptsize \centering University of California, Santa Barbara}

\teaser{
  \centering
  \includegraphics[width=\linewidth]{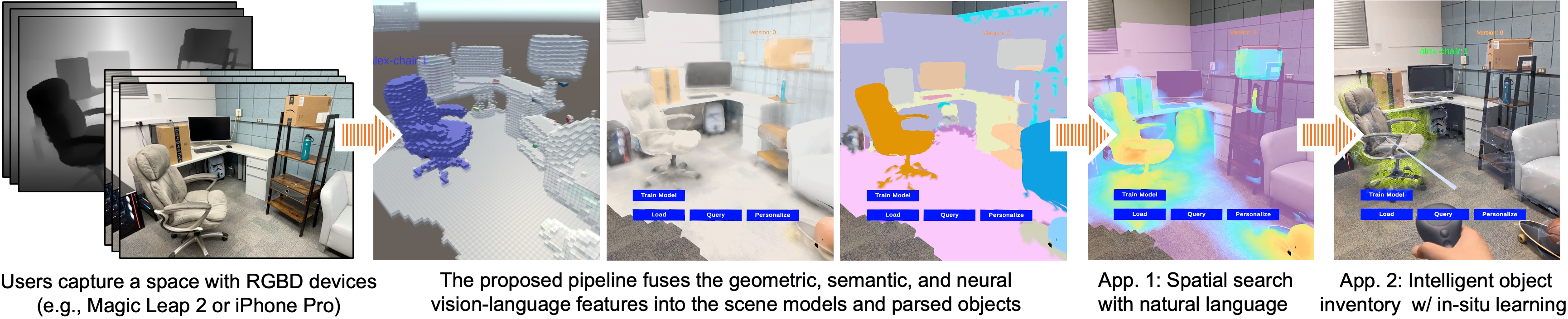}
  \caption{We propose a multimodal 3D reconstruction pipeline that prepares physical spaces for vision-language perception and object-level interactive machine learning. Within a few minutes after a user scans the environment, they can search the space with abstract natural language queries or interact with physical objects for spatially aware AI applications through novel AR interfaces.}
  \label{fig:teaser}
}

\abstract{
Seamless integration of virtual and physical worlds in augmented reality benefits from the system semantically ``understanding'' the physical environment. AR research has long focused on the potential of context awareness, demonstrating novel capabilities that leverage the semantics in the 3D environment for various object-level interactions. Meanwhile, the computer vision community has made leaps in neural vision-language understanding to enhance environment perception for autonomous tasks. In this work, we introduce a multimodal 3D object representation that unifies both semantic and linguistic knowledge with the geometric representation, enabling user-guided machine learning involving physical objects. We first present a fast multimodal 3D reconstruction pipeline that brings linguistic understanding to AR by fusing CLIP vision-language features into the environment and object models. We then propose ``in-situ'' machine learning, which, in conjunction with the multimodal representation, enables new tools and interfaces for users to interact with physical spaces and objects in a spatially and linguistically meaningful manner. We demonstrate the usefulness of the proposed system through two real-world AR applications on Magic Leap 2: a) spatial search in physical environments with natural language and b) an intelligent inventory system that tracks object changes over time. We also make our full implementation and demo data available at (\url{https://github.com/cy-xu/spatially_aware_AI}) to encourage further exploration and research in spatially aware AI.} 

\keywords{Augmented reality, artificial intelligence, interactive machine learning, scene understanding}

\begin{document}

\definecolor{code_bg}{HTML}{0D089F}


\firstsection{Introduction}

\maketitle

3D scene understanding of physical environments is crucial for context-aware augmented reality (AR). Research and industry endeavors have been steadily advancing the sensing and sensemaking capabilities of mobile computational platforms~\cite{feiner1997touring,tran2023wearable}. 
Modeling and understanding basic geometric configurations such as room size, solid surfaces, and occlusions enable realistic virtual content placement and interactions \cite{gal2014flare, nuernberger2016snap, spatialmapping}. A good semantic understanding and 3D segmentation can reveal the what and where of common objects in an environment to enable complex interactions and deeper blending of the virtual with the physical \cite{hettiarachchi_16_annexing, yue_17_scenectrl, lindlbauer2018remix, tahara_2020_retargetable_ar, suzuki2020realitysketch, kari_23_scene_responsiveness}. Leveraging the power of recent large multimodal models (LMMs) and large language models (LLMs), we can even perform simple spatial and linguistic reasoning in complex real-world scenes \cite{kerr2023lerf,sharma_octo_2024}.


If we continue to push the envelope, what new forms of scene understanding and reasoning are in store for context-aware AR and its applications? Inspired by the unified visual and linguistic knowledge latent embeddings from recent neural vision-language models (e.g., OpenCLIP~\cite{cherti2023reproducible}) that enabled unprecedented world-sensing capabilities. We believe a promising holistic approach to probe this direction is to build \textbf{a multimodal fusion pipeline that integrates the 1) geometric, 2) semantic, and 3) linguistic (vision-language) information of real-world scenes and objects into a unified 3D representation.} To this end, we implemented a TSDF-based \cite{curless1996volumetric} 3D reconstruction and segmentation pipeline that fuses the deep linguistic features from RGB frames into the 3D representations of the physical space and individual objects. 

The 3D fusion of neural vision-language features automatically enables linguistically meaningful spatial computing. While previous AR spatial search tasks are constrained by limitations of close-set detection or segmentation models, it is now possible to search for arbitrary objects, or even respond to abstract natural language queries in a physical space. We show in \Cref{fig:teaser} and \Cref{fig:application_results}~Application 1 that a heat map responding to the query ``Things that might be dangerous to babies'' highlights the most probable areas through the AR headset. The physical environment, imbued with vision-language features, can provide valuable information about itself via AR interfaces.     


We fuse context-relevant vision-language features into the 3D scene at the abstraction levels of scene voxel, vertex, and individually segmented physical object. Such CLIP-embedding-fused and semantically-indexed objects are more intelligent ``virtual twins'' than previous 3D geometric models. They become even more powerful when we add user-in-the-loop interactive machine learning controlled by AR interfaces. The user's interactions with physical objects provide valuable model steering instructions to train a personalized machine learning model, e.g. for intelligent inventory management.


A good use case for demonstrating the advantages of pairing the proposed multimodal 3D fusion with a user-guided machine learning mechanism, which we dubbed ``in-situ" learning, is to track the changes of physical objects in the real world. Unlike a conventional version control system (e.g., Git~\cite{git}) that can compare the difference between two saved states, if we move a specific red coffee mug from one desk to another, the simple spatial translation alters an object's orientation or volumetric representation, yet no change would occur semantically. In order to provide users with useful information, the item needs to be identified as the same entity. In office spaces where shared objects rarely stay at the same place or orientation, naive mesh comparison only produces noise. What is needed is a user-trainable classifier that can learn to remember arbitrary physical objects and quickly optimizes for users' changing needs and the task at hand.

To this end, we present a proof-of-concept intelligent object inventory system, demonstrating the ability to remember and re-identify objects in physical environments enabled by multimodal 3D fusion and in-situ learning. The demos run on Magic Leap 2 and work in complex real-world spaces. Once trained with simple user guidance, the system can reveal objects that are missing or remain unchanged over time in a tracked space. In \Cref{fig:application_results}~Application~2, we show that when a colleague's rolling chair was removed from the tracked scene, we can travel back in time to reveal the disappeared chair at its previously recorded location. The accompanying video reveals more object changes in this scene.

Our contributions can be summarized as follows:
\begin{itemize}
  \item We present a custom multimodal 3D reconstruction workflow that fuses both the semantic and neural vision-language features into the 3D models of the environment and objects, unlocking novel context-aware AR interfaces for physical spaces and objects.
  \item We demonstrate the enhanced effectiveness of the fusion pipeline when coupled with in-situ learning in real-world spaces with two novel AR applications on Magic Leap 2: a)~spatial search with natural language and b)~an intelligent inventory prototype that can track physical object changes.
  \item We share system design details and open source our implementation and demo dataset to help fellow researchers develop future spatially aware AI applications based on our system.
\end{itemize}

This paper is organized as follows: \Cref{sec:related} discusses the related work. \Cref{sec:system} describes the design decisions and technical details of the fusion pipeline, in-situ machine learning, and the scene manager. \Cref{sec:demos} demonstrates two real-world application scenarios of the proposed system. We conclude the paper by discussing our vision, the limitations, and future work in \Cref{sec:discuss}.

\section{Related Work}\label{sec:related}
Our work is broadly inspired and germane to topics in mixed reality, computer vision, machine learning, and HCI. In this section, we discuss related work in the areas of AR Scene Understanding, AR Scene Authoring, Physical Interaction in AR, Interactive and Online Machine Learning, Open-Vocabulary 3D Perception, and Version Control for Non-traditional Media.

\subsection{Augmented Reality Scene Understanding}\label{sec:scene_understanding}

Scene understanding gives semantic meaning to reconstructed 3D models of the physical environment. This allows AR headsets to know not just the geometry but also the what and where of objects in the space, which is needed to unlock context-aware AR and interactions. SLAM++~\cite{slas_2013_slam++} generates an object-level scene description relying on prior knowledge. However, the requirement of a library of known objects prevents this system from generalizing to arbitrary scenes. 
FLARE~\cite{gal2014flare} creates AR object layouts that are consistent with the geometry of the physical environment. SnapToReality~\cite{nuernberger2016snap} helps align virtual content to real-world 3D edges and surfaces. Spatial mapping in Hololens~\cite{spatialmapping} can infer semantic surfaces such as walls, floors, platforms, and ceilings.

Recent leaps in computer vision, foundation models, large language models (LLMs), and large multimodal models (LMMs) provide new directions in tackling object-level scene understanding tasks. Chen et al.~\cite{chen_context-aware_2020}, PanopticFusion~\cite{narita_panopticfusion_2019}, and Panoptic Multi-TSDFs~\cite{schmid_panoptic_2022} go beyond geometric-based AR by combining semantic segmentation with dense 3D reconstruction to achieve object-level 3D understanding, removing some of the constraints that SLAM++ was beholden to. Previous work has also explored characterizing scene context (such as location and orientation of objects) to support virtual content placement and realistic interactions with the physical environment~\cite{tahara_2020_retargetable_ar}, and automatic annotation of unprepared environments using machine learning to detect, recognize and segment objects intelligently~\cite{pucihar2023fuse}.

In a more challenging open-vocabulary setting, Yoffe and Sharma proposed OCTOPUS and OCTO+~\cite{yoffe_octopus_2023,sharma_octo_2024} to automatically place arbitrary objects on the most suitable surface in AR. They chained a series of state-of-the-art ML methods to build a Mixture of Experts System, and used Segment Anything Model (SAM)~\cite{kirillov2023segany} to identify individual objects, CLIP and clip-text-decoder~\cite{radford2021clip, odom_22_clipdecoder} to generate object labels, and ViLT~\cite{kim_vilt_2021}, CLIPSeg~\cite{lueddecke22_cvpr}, and Grounding DINO~\cite{liu_grounding_2023} to verify object guesses. Like many mixture of experts systems, they used LLMs or LMMs such as  GPT-4, GPT-4V~\cite{openai2024gpt}, and LLaVA~\cite{liu2023improvedllava} as the ``brain'' to reason about appropriate locations for object placement based on curated text and image inputs from various upstream models.

Compared to their work, which focuses entirely on the question of how to place content in 2D image frame observations of 3D scenes, we tackle the much more general problem of embedding the semantic features within the 3D geometric representation, enabling additional levels of spatial reasoning.

\subsection{Augmented Reality Scene Authoring}

Authoring tools complement automatic scene understanding by allowing content creators or end users to assign semantic meaning or information to the reconstructed scene even in the absence of fully reliable automatic detection and segmentation. Early authoring systems such as Columbia's ``touring machine'' \cite{feiner1997touring} relied on designers to manually attach information to the environment, whereas the follow-up MARS system utilized offline and online authoring tools~\cite{hollerer1999mars}. Other AR applications \cite{mann1997wearable, spohrer1999worldboard} have also explored the placement of virtual information on recognized real-world objects.

Immersive authoring~\cite{lee_immersive_2005} in AR allowed users to parse objects from the reconstructed 3D model while moving around in the actual physical space. Interactive online systems such as SemanticPaint~\cite{valentin2015semanticpaint} and Semantic Paintbrush~\cite{miksik2015paintbrush} continuously learn from the user's segmentation input to predict object labels for new unseen voxels as the user captures the environment. Huynh et al. proposed In-Situ Labeling \cite{huynh_insitu_2019} to facilitate more effective language learning in AR settings. SceneCtrl~\cite{yue_17_scenectrl} and HoloLabel~\cite{agrawal2022hololabel} provide user-in-the-loop scene editing and labeling.

Unlike the above scene authoring tools that require careful interactive operations on voxels or meshes, this work has individual physical objects automatically segmented and labeled during the 3D reconstruction process. Users interact with the densely labeled scene at the object level for personalization by directly selecting individual objects.

\subsection{Interacting with Physical Objects in AR}

Going beyond geometric-based AR, recent context-aware AR focuses on novel interfaces that interact with virtual twins of physical objects in the environment, creating illusions that tightly blend the physical and the virtual. SceneCtrl~\cite{yue_17_scenectrl} and Remixed Reality~\cite{lindlbauer2018remix} enable manipulation of virtual versions of the physical environment. Annexing Reality~\cite{hettiarachchi_16_annexing} uses physical objects as proxies for virtual content to reduce the visual-haptic mismatch. RealitySketch~\cite{suzuki2020realitysketch} captures user sketchings to create virtual elements bound to physical objects, which dynamically respond to real-world changes. TransforMR~\cite{kari_21_transformr} can replace real-world humans and vehicles with pose-aware virtual object substitutions to produce semantically coherent MR scenes. Kari et al.~\cite{kari_23_scene_responsiveness} demonstrated the concept of Scene Responsiveness which maintains visuotactile consistency in situated MR through visual illusions that hide, replace, or rephysicalize real objects with virtualized objects and characters.

Our work is related to the above research as we provide novel interfaces for users to interact with the physical room, in our case through natural language and by tracking physical objects' changes over time. It differs from the above works in that we integrate deep vision-language features into the 3D models to automatically identify (segment and label) and remember individual objects for the application scenarios.


\subsection{Interactive \& Online Machine Learning for AR}
\label{sec:ml_ar}

Interactive ML and online ML are distinct learning paradigms that often go hand in hand in real-world interactive AR/MR applications. In these applications, training data becomes available in the form of a stream during the user's interaction with the environment.

To learn from previous user gestural and verbal input to predict the segmentation and object labels for new unlabeled parts of the 3D scan, SemanticPaint~\cite{valentin2015semanticpaint} proposed a streaming random forests algorithm that trains on voxel-oriented patches (VOPs), which are geometric and color features computed from raw TSDF volumes. Semantic Paintbrush \cite{miksik2015paintbrush} adopted the same VOPs as object features (RGB, surface normal vector, and 3D world coordinate) to train a similar streaming decision forest. ScalAR \cite{qian_22_scalar} also used a decision-tree-based algorithm to learn from the user input.

Unlike previous interactive AR systems that trained decision trees on low-level features, this work utilizes deep vision-language models to generate semantic and linguistically meaningful deep latent features for the environment and individual objects. We propose a multimodal representation that considers an object's geometrics (voxels), appearance (RGB), and vision-language features (CLIP) to produce meaningful object graph representations that are robust against issues that low-level features suffer from, such as changing lighting conditions, orientation, and over-simplified semantics. We also propose our own version of interactive online ML, dubbed ``In-Situ Machine Learning'' (see \Cref{sec:insitu}), to refine and improve the performance of our automatic semantic segmentation and object recognition.

\subsection{Open-Vocabulary 3D Perception}

The attribute ``open-vocabulary'' describes a system that can recognize objects matching a free-text description, which may contain arbitrary natural-language descriptors (e.g., ``a chair whose color is somewhere between blue and green'') or abstract concepts (e.g., ``what can I use to prop open a door?''). This is a much more flexible, intuitive, and ultimately more useful paradigm in many scenarios compared to the traditional computer vision approach of classifying objects into a pre-determined semantic taxonomy, which is inflexible and cannot be exhaustive.

Several works have presented open-vocabulary systems for 2D image segmentation and understanding, either at the level of patches or entire images~\cite{radford2021clip, alayrac2022flamingo, jia2021scaling}, while others have focused on dense, per-pixel representations~\cite{gu2021open, li2022language, rao2022denseclip}. These systems became possible because of the massive amount of paired images and text available on the internet that were used as training data for vision-language foundation models. In contrast, for 3D data, it is more difficult to directly develop the 3D-language connection, due to the lack of large datasets of paired geometry and text. To address this, a number of works have attempted to bootstrap 3D open-vocabulary perception by distilling or otherwise lifting 2D open-vocabulary models to operate on 3D data~\cite{kerr2023lerf, peng2023openscene, zhang2023clip, ding2023pla, yang2023regionplc}. 

Our system, specifically the {\em Spatial Search with Natural Language} feature, belongs to this latter category, lifting CLIP features into 3D by back-projecting them into a voxel grid, using a modification of the popular TSDF fusion algorithm. Our system is primarily differentiated by its design to support interactivity in AR. Most importantly, it operates with low latency, without requiring expensive components such as 3D convolutional neural networks or training neural radiance fields. This enables a smooth and familiar AR scanning workflow. In addition, our system builds an implicit surface mesh representation rather than relying on point clouds or density volumes. This is more amenable to downstream processing and rendering with the traditional graphics pipeline, meaning that it can easily be integrated into existing AR platforms and applications.

\subsection{Version Control for Novel Media}
\label{sec:vcs_related}

One of our demonstration applications, intelligent object inventory, tracks certain object changes that are akin to the behaviors in version control systems. Software developers are most familiar with text-based version control systems (VCSs) such as Git~\cite{git} that keep track of changes in source code. The research community has explored version control interfaces and techniques on novel media other than text editing. Time-Machine Computing~\cite{rekimoto_19_tmc} tracks computer desktop states and allows users to visit a previous state. MeshGit \cite{denning_meshgit_2013} proposed a mesh edit distance to measure the dissimilarity between two polygonal meshes in 3D modeling workflows. SceneGit~\cite{carra_19_scenegit} tracks object-level element changes in a 3D scene as well as finer granularity changes at the vertex and face level. The \textit{Who Put That There} system~\cite{who_put_that_there} records virtual objects' spatial trajectories from the user's direct manipulation in 3D VR scenes. 


VRGit~\cite{zhang_23_vrgit} facilitates synchronous collaboration for manipulating and comparing VR object layouts immersively. While the system provides well-defined spatial versioning features, it operates on a library of predefined models because of its VR nature. Research with physical artifacts in mind, such as Catch-Up 360~\cite{perteneder_catch-up_2015} and works by Letter et al. \cite{letter_comparing_2023} focused on the changes of a single object rather than room-size environments like in VRGit. AsyncReality~\cite{fender_causality-preserving_2022} used external devices to volumetrically capture physical events for later immersive playback.

Our work differs substantially from the above version control research in that physical objects change in ways different from source code or 3D models -- spatial translation, deformation of non-rigid objects, and appearance changes based on lighting conditions. Naive comparison between two mesh models only introduces counterproductive noises, even if they are perfectly aligned. This work, however, maintains object identity by relying on deep vision-language features embedded in the reconstructed 3D environment. We demonstrate one of the first intelligent inventory systems that automatically track basic object changes (missing/unchanged) in real-world environments in \Cref{sec:inventory}.

\section{System Overview}\label{sec:system}

At the heart of our spatially aware AI system that informs the AR user interfaces in this paper is a custom 3D reconstruction pipeline with integrated vision-language fusion and 3D segmentation workflow. The pipeline starts with capturing a physical space -- a user walks around with an RGBD device that captures registered RGB images and depth maps to reconstruct the 3D model with geometric, semantic, and linguistic understanding (see \Cref{fig:teaser}).  \Cref{fig:system} shows the system overview. The accompanying video also demonstrates the interactive possibilities and flow of the system. We will discuss the design of the main components in this section.

\subsection{Multimodal 3D Scene Model Fusion}
\label{sec:integration}

Our multimodal scene volume is represented by a multi-channel voxel grid defined over the scene. The channels of this volume are organized into three components: geometry, language, and semantics. The geometric component is a single-channel TSDF volume produced using the typical TSDF fusion algorithm \cite{curless1996volumetric} according to the following running-average update rule:
\begin{equation}
    D_{i+1}(x) = \frac{D_i(x)W_i(x) + d_{i+1}(x)w_{i+1}(x)}{W_i(x) + w_{i+1}(x)},
\end{equation}
where $D_i(x)$ is the accumulated TSDF estimate over all past views for voxel $x$ at time $i$, and $d_i(x)$ is the TSDF estimate for voxel $x$ from the current view at time $i$. $w$ represents a per-view weight, and $W$ is the total accumulated weight (we refer the reader to Curless \& Levoy \cite{curless1996volumetric} for further details).

We then propose a simple mechanism to extend the scene volume with additional channels, which are populated by fusing feature vectors from image-aligned 2D feature maps as follows:
\begin{equation}
    F_{i+1}(x) = \frac{F_i(x)W_i(x) + f_{i+1}(x)w_{i+1}(x)}{W_i(x) + w_{i+1}(x)},
\end{equation}
where $f_{i}(x)$ is a feature vector sampled from view $i$ by perspective projection from voxel $x$, and $F$ is the generated multi-channel feature volume. The main advantage of fusing features in this manner is that by averaging across views, we develop a more accurate multi-view feature and label estimate over time.

We leverage this extension to build the semantic and language components of the volume by fusing in two additional sets of 2D feature maps. The first one (object semantics) is a per-pixel class probability distribution, computed using the panoptic segmentation from k-means Mask Transformer \cite{kmax_deeplab_2022}. The second one (language) is a per-pixel CLIP feature computed using OpenCLIP \cite{cherti2023reproducible}. Since CLIP's feature output for a given image is only a single feature vector with no spatial dimensions, we tile each image into overlapping patches to produce a coarse 2D CLIP feature map. We then define a continuous CLIP feature across the image using bilinear interpolation.  While the parameters can vary among capture devices and scenes, our setup resizes input frames to $1024\times768$ px and uses $256^2$ px patches with a stride of 128 px.

\begin{figure}[t!]
  \centering \includegraphics[width=\linewidth]{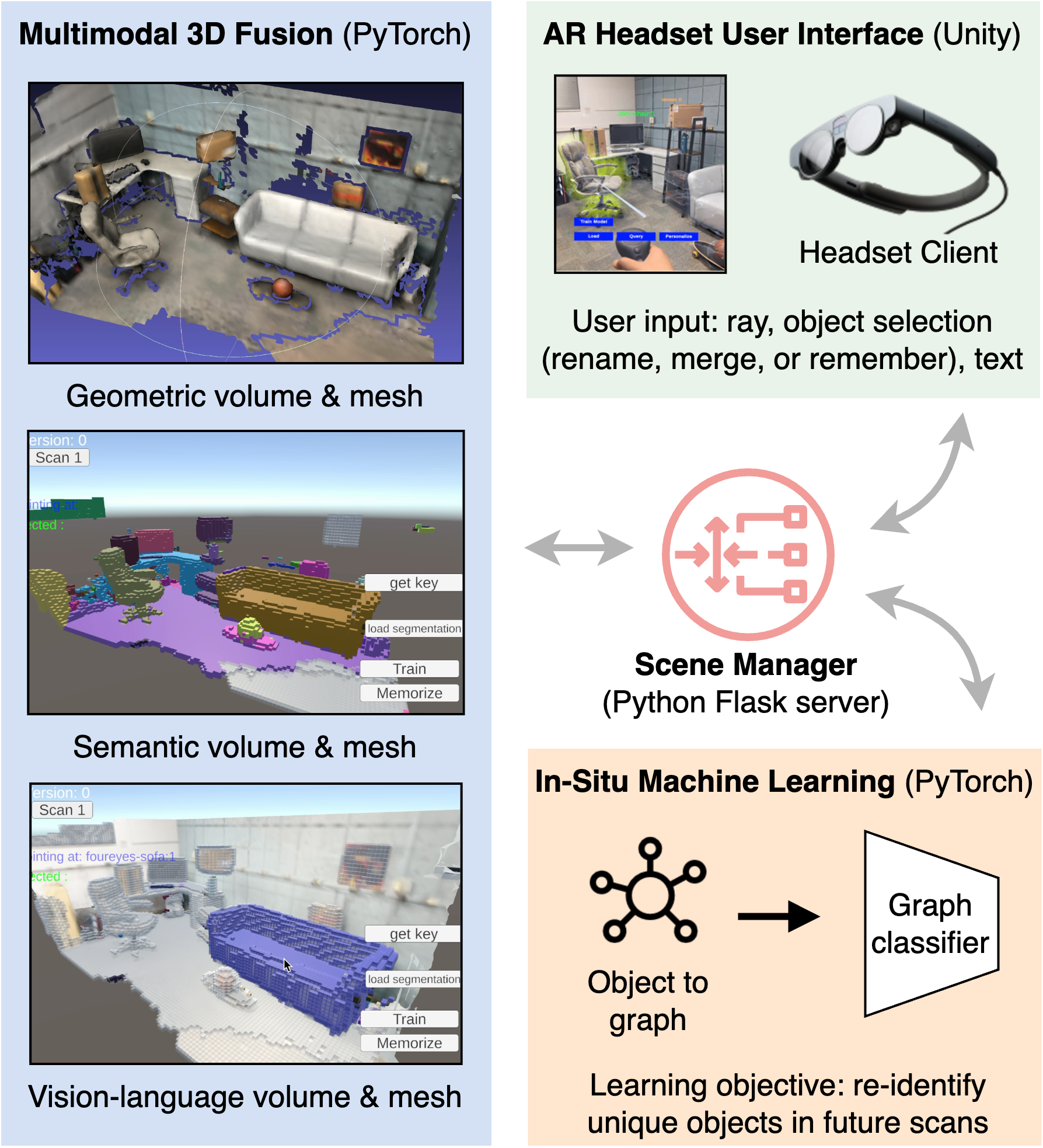}
  \caption{System overview.}
  \vspace{-0.3cm}
  \label{fig:system}
\end{figure}

Finally, the fusion process results in a per-voxel TSDF estimate, class probability distribution, and CLIP feature, that we use to support downstream applications (\Cref{fig:system} left). This process exhibits two properties that make it highly amenable to AR applications: 1)~all three volume components are constructed using a running average update rule, so the process is incremental and can accept new input views at any time without needing to revisit earlier views; 2)~no iterative optimization is required, leading to fast online reconstruction.

\subsection{Post-Processing and Scene Manager}
\label{sec:postprocessing}

\begin{table*}[t]
\centering{\caption{\label{table_performance}Post-processing performance with different scenes and voxel size. The cluttered ``Day 1" and simpler ``Day 2" scenes presented in the paper (\Cref{fig:system}~and~\Cref{fig:application_results}) and accompanying video adopted 4cm voxels. Text query ``things that might be dangerous to babies" was used to test query response time. The ``N/A" in the 2cm setting is due to insufficient GPU memory.}}
\begin{tabular}{ |c|c|c|c|c|c|c|c|c| } 
\hline
Scene & Scanned & Voxel Size & Voxel Grid & Semantic & Post-Processing & Text Query & Database & GPU Memory \\
& Views & (cm) & (x,y,z) & Segments & (seconds) & (seconds) & (MB) & (MB) \\
\hline
& & 2 & $251\times100\times136$ & 3,319 & $\sim168$ & N/A & 7,367 & $\sim22,827$ \\ 
\multirow{2}{5em}{\centering ``Day 1"} & \multirow{2}{3em}{\centering 213} & 4 & $128\times53\times71$ & 617 & $\sim85$ & $\sim1.26$ & 1,073 & $\sim6,980$ \\
& & 8 & $67\times30\times38$ & 161 & $\sim77$ & $\sim0.23$ & 175 & $\sim4,576$ \\ 
& & 16 & $37\times18\times22$ & 51 & $\sim76$ & $\sim0.03$ & 34 & $\sim4,400$ \\ 
\hline
``Day 2" & 121 & 4 & $128\times54\times80$ & 404 & $\sim52$ & $\sim0.93$ & 1,211 & $\sim7,359$ \\ 
\hline
Large Office & 103 & 8 & $172\times33\times117$ & 616 & $\sim51$ & $\sim1.70$ & 1,491 & $\sim17,050$ \\ 
\hline
\end{tabular}
\vspace{-0.3cm}
\end{table*}

Following the multimodal 3D fusion, we perform three post-processing tasks to support downstream applications.

\textbf{1) Mesh extraction.} We run Marching Cubes \cite{lorensen1998marching} to extract a triangle mesh from the TSDF volume. This allows for convenient rendering and integration with existing AR graphics pipelines. \Cref{fig:teaser} shows various mesh visualizations rendered in AR headsets.

\textbf{2) 3D semantic segmentation.} To create useful spatial awareness for AR, we are interested in going beyond per-voxel semantic information to delineate full objects that users can more easily select and manipulate. We therefore build on the class probability volume developed in \Cref{sec:integration} by first labeling each voxel with the class for which it has the highest predicted probability, and then segmenting consecutive volumes according to those labels using a custom 3D flood fill implementation. Similar to classic 2D flood-fill algorithms that find connected regions on images \cite{tint_fill}, our 3D method clusters voxels of the same segmentation class in the 3D volume grid to parse individual objects in the user's surroundings, extracting the complete object boundary, shape, and identity, with no user intervention required (see \Cref{fig:system} semantic volume).

\textbf{3) Intelligent object inventory.} During the object parsing process, the Scene Manager creates an object inventory by associating the per-voxel CLIP features with the individual objects. The scene manager, as shown in \Cref{fig:system}, is the central communication hub that a)~manages multiple versions of environment models, b)~sends and receives data between the AR user interface via HTTP requests, and c)~utilizes the in-situ machine learning to ``remember and re-identify" unique objects for the intelligent object inventory. 

Compared to conventional 3D reconstruction focusing on user manipulation of the object model's geometric representations (e.g., mesh or volume), \textbf{the proposed multimodal fusion associates each object's semantic/identity, metadata, and vision-language (CLIP) features with its geometrics, producing intelligent virtual twins}. With CLIP features attached to objects, novel spatially-aware-AI interfaces are unlocked through interactive machine learning, which we will discuss in the next section. After these post-processing steps, users can easily interact with physical objects through the AR virtual pointer (see \Cref{fig:system} user input).

\textbf{System Performance.} We host our system on a local server with a single NVIDIA RTX 3090 GPU (24GB of VRAM) to perform multimodal fusion and in-situ learning. The voxel size used in TSDF reconstruction is a key parameter that affects both the fusion quality and the object inventory. We present the post-processing performance in \Cref{table_performance} for the two office test scenes shown in our paper and the accompanying video. A small voxel size like 2cm reconstructs the space at a high definition but comes at a much higher computation cost: a $5m\times3m$ space barely fits into the VRAM, yet the text query task crashes due to insufficient memory. A large voxel size of 16cm reduces the memory footprint and the processing time, but the ``bulky" 3D model failed to reconstruct or segment many small objects, making it a better choice for outdoor scenes.

We find the 4cm voxel, about the volume of a computer mouse, strikes a good balance between computation, memory, and the smallest detectable objects in cluttered scenes. To test even larger real-world scenes like our 124 $m^2$ (1,300 $ft^2$) office space that seats 14 people, we found 8cm voxels sufficient to track common objects (last row in \Cref{table_performance}). In short, the GPU acceleration can make the demo scene ready for user interaction in just two minutes with our current system design and hardware, but the limited VRAM becomes the bottleneck for tracking large areas with high precision. Depending on the use case, this system can be scalable with 1)~more GPUs, 2)~half-precision tensors for a smaller memory footprint, or 3)~a CPU-only setup when response time is not the priority (e.g., daily inventory tracking).

\subsection{“In-Situ” Machine Learning}
\label{sec:insitu}

We propose a novel interactive and online machine learning concept called ``In-Situ'' Learning to improve AR experience in complex real-world environments. Since we have attached rich multimodal features to individual objects, one practical optimization objective for the in-situ learning model is to learn to remember and re-identify individual objects across different scans. \textbf{We define in-situ learning as the process of encoding real-time data into a neural network, such that the network itself serves as both the knowledge container and decision-making unit for downstream tasks}, e.g., as a probe to identify changes such as new or missing objects. This is related to AR works discussed in \Cref{sec:ml_ar} and works that perform online neural scene encoding (Feng et al. NARUTO \cite{feng2024naruto}, Sandstrom et al. Point-SLAM \cite{feng2024naruto}), but we introduce the additional temporal dimension to enable object tracking across multiple time points (room scans on different days) in the second application scenario.

The ``in-situ'' (Latin for ``in position'' or ``on site.'') nature of our learning concept is characterized by the following observations:

\textbf{1)} As we live in a complex and constantly evolving world, the neural vision-language features that represent objects and contexts also change dynamically. These \textbf{unlabeled training data} are only generated at the moment when the user captures the physical space.

\textbf{2)} Novel \textbf{ground truth samples} that guide the supervised learning (e.g., personalized object names and merged segments) are generated only when the user interacts with the environment. Unlike offline-collected large-scale datasets that present a universally accepted ground truth of our world~\cite{deng_imagenet_2009, lin_microsoft_2014}, the ground truth in in-situ learning varies among users and over time.
  
\textbf{3)} Similar to typical interactive machine learning, the \textbf{model's performance is evaluated by the user} instead of by a fixed benchmark. It's at the user's discretion to decide if the model is sufficiently optimized, or else they can provide more training data by re-scanning the space or re-labeling incorrectly classified objects to fine-tune the model.

\textbf{4)} Also similar to interactive machine learning, the model always immediately reflects \textbf{individual user's latest annotations and preferences}, which is in contrast to the typical batch incorporation of user feedback necessitated by large offline-trained models. 

\begin{figure}[t]
  \centering \includegraphics[width=0.7\linewidth]{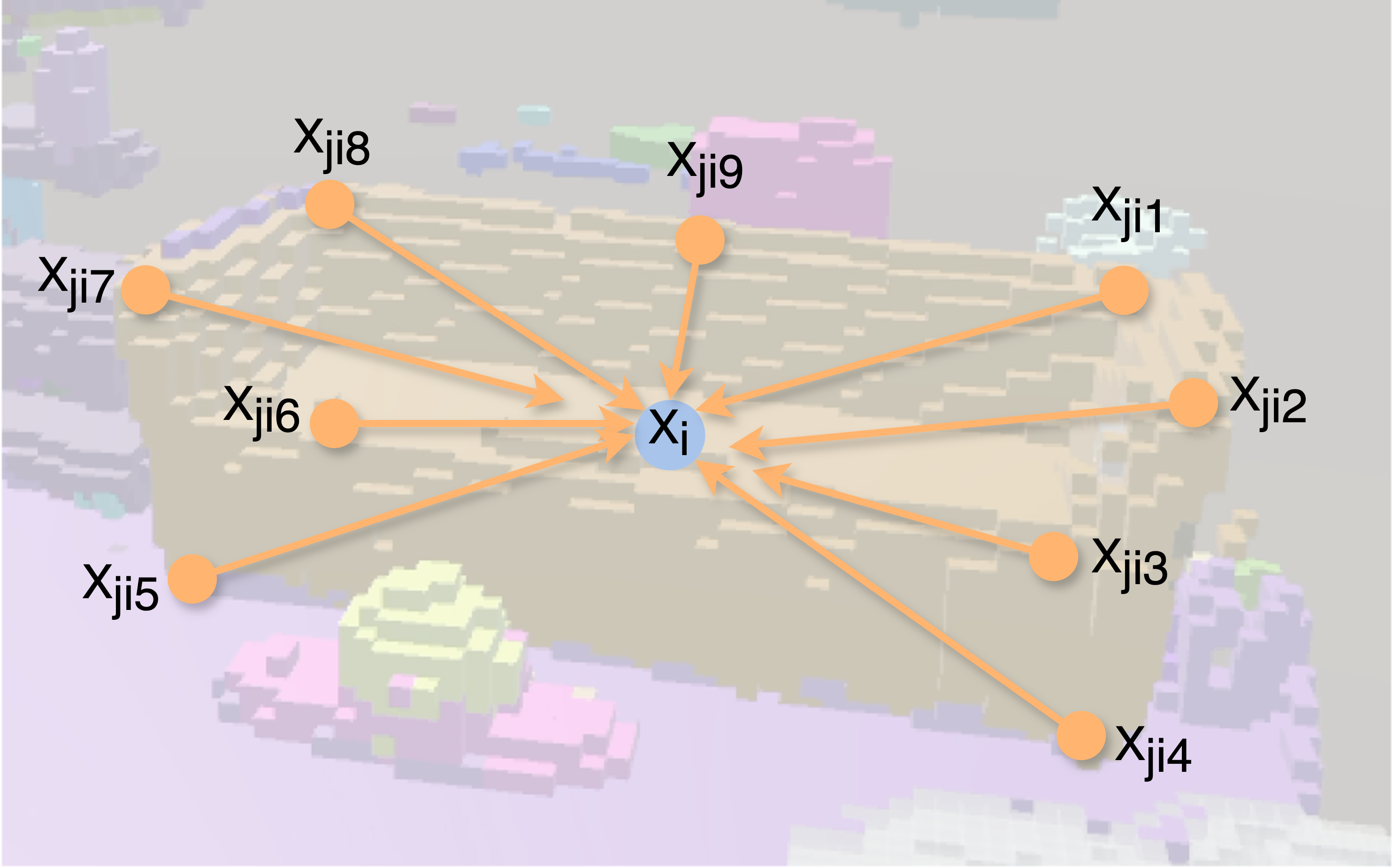}
  \caption{During the real-time in-situ training, we sample a sparse graph from an object's voxel representation stochastically, with the voxel location's CLIP feature as the node attribute. This design choice converts the challenging irregular 3D object classification problem into a simpler graph classification problem, which enables us to identify physical objects across multiple scans of the space. The sofa graph above is oversimplified for visualization purposes.}
  \vspace{-0.3cm}
  \label{fig:sofa_graph}
\end{figure}

To accommodate the task of tracking physical objects, we convert an object's irregularly shaped volume representation to a graph representation to optimize for online machine learning. Unlike previous scene authoring AR work that trained on low-level TSDF features alone~\cite{valentin2015semanticpaint, miksik2015paintbrush, qian_22_scalar}, the multimodal intelligent virtual twins allow us to create a novel graph representation that combines the geometric, semantic, and linguistic features for every object in the scene. As shown in \Cref{fig:sofa_graph}, we treat every voxel as a node pointing to the object's centroid, which converts an object's irregular voxel representation into a dense graph representation. For efficiency and data augmentation, we stochastically sample 30 voxels from the dense representation in each training iteration to generate a sparse graph, whose node attribute is the voxel location's OpenCLIP~\cite{cherti2023reproducible} vision-language feature, which we found sufficient to re-identify objects and reveal object changes (\Cref{sec:inventory}) without having to align the scene models for naive mesh comparison. Additional properties, such as RGB values, the geometric 3D vector pointing at the object's centroid, and relative spatial relationships to other objects can also be integrated as node or edge attributes based on specific task needs.


In other words, we turn a hard 3D object classification task into an easier graph classification task, which maintains its effectiveness even if the object or the environment changes dynamically (spatial translation, non-rigid deformation, varying lighting conditions). Additionally, the in-situ model is incrementally fine-tuned as the user provides new inputs from subsequent scans. Specifically, to learn the graph-based objects, we adopt a dynamic graph CNN~\cite{wang2019dynamic} as the backbone of the in-situ model to train a graph classifier that predicts if the graph belongs to a class label previously trained on, or an unknown class (e.g., background objects not marked by user). A single in-situ model is trained for a specific space -- it can be tailored by one user for personalization or shared between a group of users for collaboration and information exchange.

To summarize, in-situ learning's novelty lies in the ``just-in-time'' user-generated data and the evaluation metric that is based on user satisfaction. In real-world applications, system designers should choose the specific type of machine learning paradigm (e.g., supervised or self-supervised), the model architecture, and the training strategy that best supports the task at hand.

\begin{figure*}[ht!]
  \centering \includegraphics[width = \textwidth]{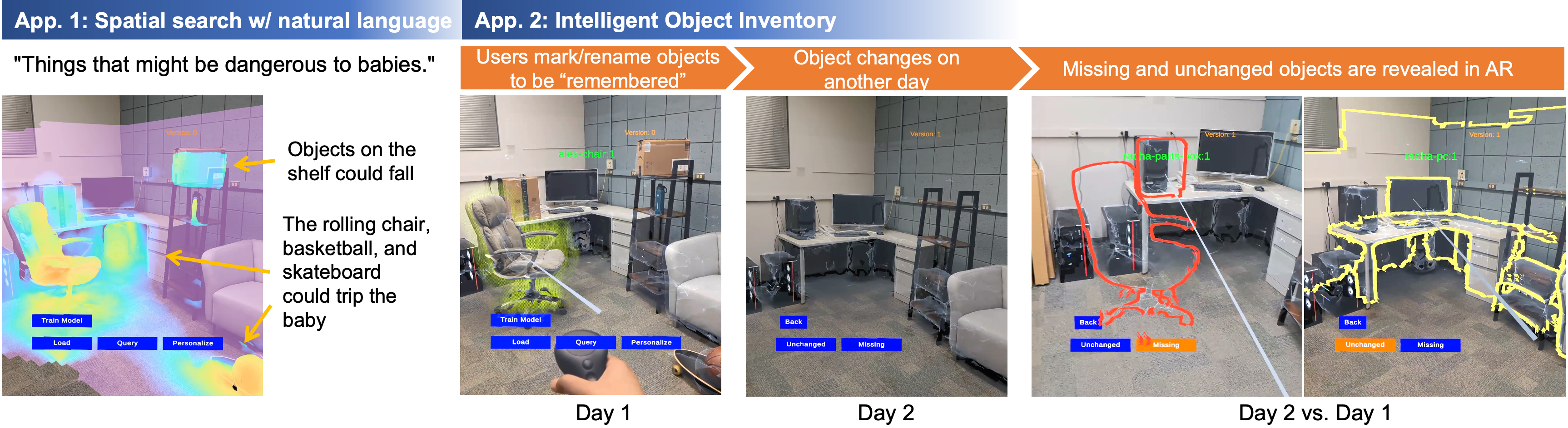}
  \caption{Two prototype applications developed on Magic Leap 2 AR headset, demonstrating the potential of the proposed multimodal 3D fusion pipeline and ``in-situ'' machine learning for real-world scenarios.}
  \vspace{-0.3cm}
  \label{fig:application_results}
\end{figure*}

\section{Prototype Applications}\label{sec:demos}

In this section, we showcase two real-world Magic Leap 2 AR prototype applications to demonstrate the potential of the proposed multimodal 3D fusion pipeline and new tools and interfaces for users to interact with physical spaces and objects when used in conjunction with ``in-situ'' machine learning. 

\subsection{Spatial Search with Natural Language}\label{sec:query}

\begin{figure*}[ht!]
  \centering \includegraphics[width = \textwidth]{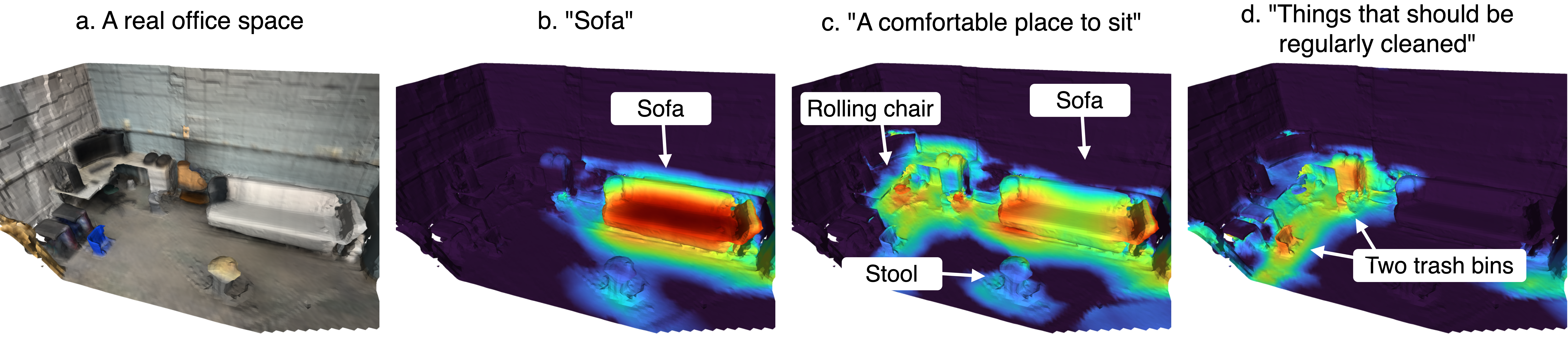}
  \vspace{-0.5cm}
  \caption{From object names to abstract natural-language queries, we show more examples of spatial search in a real-world environment.}
  \vspace{-0.5cm}
  \label{fig:queries}
\end{figure*}

We foresee a future of sophisticated real-virtual interactions, requiring deep understanding that goes beyond discrete objects with labels. By fusing 3D vision-language features into the 3D models, our system permits spatial search in a physical space using natural language. The user may issue complex, even abstract queries, for example, ``things that might be dangerous to babies.'' The system responds by highlighting matching regions in the user's surroundings to reveal potential falling hazards and objects that could trip a baby, as shown in \Cref{fig:application_results} (Application 1). \Cref{fig:queries} illustrates more example queries and results.

This search capability is built on the language component of our scene volume, composed of a multi-channel CLIP feature volume. CLIP is key to this process as it embeds images and text into a shared feature space. Thus we can cast natural language search as building a map over the scene of the similarity between scene CLIP features and the CLIP embedding of a user-supplied text query. To enable this, we first resample the CLIP feature grid using trilinear interpolation to obtain a CLIP feature at each mesh vertex. We then compute the similarity of each vertex feature to the query feature, relative to a set of negative queries, following the search method of CLIP Surgery~\cite{li2023clipsurgery}. However, CLIP Surgery uses a long, fixed list of negative queries to identify redundant features that come at the cost of longer computation time and higher memory load, which exceed the acceptable limits for real-time interactive AR applications. We build the negative query list as the union of all class names extracted by our 3D semantic segmentation step. Our list is therefore shorter and more relevant, allowing us to produce heatmap outputs to the query similarity efficiently while filtering out noisy responses across the scene.

Our Magic Leap 2 prototype application, as demonstrated in the accompanying video and \Cref{fig:application_results}, overlays the response heatmap on top of the environment and physical objects via the optical-see-through display to provide an immersive user experience. Leveraging this spatial search ability in AR applications can provide users with an enhanced understanding and navigation of unfamiliar spaces. It can also enable them to explore complex environments faster than they would manually.

\subsection{Intelligent Object Inventory}
\label{sec:inventory}

We imagine an intelligent AI AR companion that keeps a ``temporal and spatial inventory'' of objects for real-world environments, helping users keep track of the objects in their space. Integrating geometric and semantic knowledge into the joint 3D space makes it possible to automatically parse individual 3D objects from the environment for an object-centric user interface. As shown in our accompanying video and \Cref{fig:application_results} Day 1, the ``magical'' instantaneous highlighting and selection of physical objects in optical-see-through AR displays from any viewpoint provides an intuitive and direct interface for users to edit or personalize their space. 

Our main goal with this application is to show that when coupled with in-situ learning, the multimodal-feature-fused environment models can unlock novel spatially-aware AI user interfaces. For instance, having access to intelligent virtual twins of every physical object in the room makes it feasible to train a machine-learning model to ``learn and remember'' physical objects, maintain object identities, and track object changes without aligning any noisy unstructured mesh models. To this end, we introduce a basic intelligent inventory system to
visually present one interpretation of object changes in real-world environments. 

We briefly discussed in \Cref{sec:vcs_related} that the concept of ``changes'' from text-based version control systems does not automatically translate to the spatial, morphological, or appearance changes in physical objects. While the ``true removal'' of an object or a paragraph of text is similar, naively comparing different scans of a space captured on different days yields counterproductive noise and does not maintain object identities.  The ability to re-identify an arbitrary physical object is critical for effective user assistance. Learning to remember and re-identify an object relies on the in-situ learning model, which we introduced in \Cref{sec:insitu}. We will now discuss how it is used to reveal unchanged or missing objects in AR. This proof-of-concept AR demonstration does not yet constitute a full inventory system. Limitations are discussed in \Cref{sec:limits}.

We offer three actions 
to collect user input to determine which objects to track and train on:

1) \textit{Merge.} Users can merge multiple mesh segments into one if a single object was fragmented during the 3D reconstruction process, e.g., false boundaries introduced by shadows. This change is picked up by in-situ learning, leading to the recognition of segmented parts as the same object in future scans of this space.

2) \textit{Rename.} Users can also customize the automatically generated object labels (e.g., ``bottle:2'' to ``Joe's thermos'') to improve their utility. This feature proves particularly beneficial in collaborative environments, such as a shared office, where it can help specify the ownership of items more clearly. In collaborative settings where multiple users might adjust the same space at different times, user-specified labels naturally reduce confusion. The accompanying video showcases an actual office setting in AR, where objects are distinctly tagged with their respective owner's names. 

3) \textit{Remember.} Users can direct the system to track certain objects in the environment without further editing actions. The same objects should be re-identified with their current properties. This design provides a quick and easy way to collect ``positive samples'' that will be used to optimize the in-situ learning model for object classification.

The user's object-level personalization input provides the ground truth to guide the learning objective. The in-situ learning model is trained to classify arbitrary objects based on their neural features from randomly sampled sparse graphs, which improves robustness across different scans and avoids overfitting. Specifically, objects that are merged, renamed, or remembered are flagged as ``positive" samples with a unique label and assigned a class index in the classifier's ground truth. Through various design experiments, we arrived at a training strategy that classifies all other ``non-personalized" objects as a ``null" class with an index of zero. We sample null class features from all other voxels that do not belong to any of the personalized objects. While it performs robustly for re-identifying objects, this strategy limits the ability to differentiate newly introduced objects from the null objects, which we will discuss in \Cref{sec:limits}. The user triggers the training after they finish personalization. The training stops automatically after the model reaches its peak accuracy (over 95\% in our demo scene) plus certain cool-down epochs. Other strategies, such as adaptive learning rate, can also be used when users label new objects and fine-tune a trained model.  In our office demo scene, we set the cool-down epochs to 10 and the total in-situ model training takes less than 8 seconds.

\begin{figure}[t!]
  \centering 
  \includegraphics[width = 0.9\linewidth]{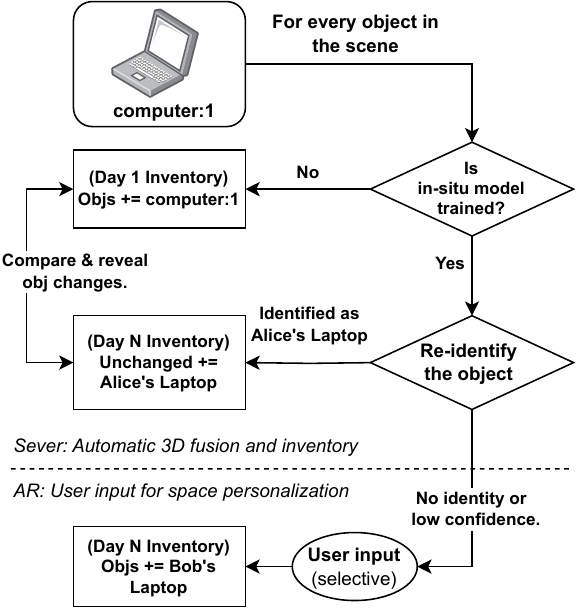}
  \caption{A flowchart describing how the Scene Manager and the user can build an intelligent object inventory with the in-situ learning model. After multimodal 3D fusion and post-processing, individual objects are passed through the in-situ model to re-identify previously existing objects and eventually reveal missing objects.}
  \vspace{-0.3cm}
  \label{fig:obj_learn_pseudocode}
\end{figure}

In \Cref{fig:obj_learn_pseudocode}, we describe how objects in a tracked space are re-identified. When a previously optimized in-situ model is available, new semantically parsed voxel clusters (object segments or objects) from the new 3D scan are first converted into a graph representation and then sent to the in-situ model for classification, i.e, to check if it matches any object that the user previously marked to remember. The object inventories of successive scenes across different days are compared to analyze which objects have been removed or remained in space.

As we show in \Cref{fig:application_results} Day 2 vs. Day 1, scans from two different days of a tracked scene are akin to \textit{``git commit''} states. The timestamps and chronology of the scans naturally form a version history. To easily inspect this history, our system renders a \textit{``volumetric diff''}, visualizing the object inventory states over time: \\
\indent 1) \textit{Unchanged objects.} Highlighting objects that were previously edited or marked by the user and can still be found in the current version of the environment. Additionally, objects that were previously renamed can still be recognized with their personalized names, even if they moved in space. \\
\indent 2) \textit{Missing objects.} Revealing all objects that were present in the previous scan of the space but are now missing from the current version of the environment. Visualizing missing objects in red hollow contours in the current space lends users additional temporal awareness of their space. In the future, this new time dimension could be the substrate for novel and more complex AR interactions.

Critically, we produce the volumetric diff unlike any other methods discussed in \Cref{sec:vcs_related}. It is the automatically segmented objects and their vision-language features that are being compared by a neural network rather than versions of the reconstructed 3D model -- misalignment in object or environment models will produce counterproductive noise instead of useful object tracking. In this proof-of-concept implementation, we manually aligned various scans to the physical room solely for the unchanged/missing objects' contours visualization. Re-identifying objects and listing which ones are unchanged or missing require no spatial alignment, which is not the focus of this work and is solvable through fiducial marker tracking algorithms (ARTag~\cite{artag_fiala}) or model matching methods (3DMatch~\cite{zeng20163dmatch}).

\section{Discussion}\label{sec:discuss}

\textbf{More use cases.} The ability to search in a physical space with natural-language queries has the obvious use case of quickly finding objects based on their description, but it can also be used creatively, leading to emergent utility, such as inspecting properties for damage or hazards, requesting decor advice, or mapping the layout of specific task-relevant items. Furthermore, this capability can serve as the backbone of future context-aware interaction systems, by identifying the virtual manipulations that are appropriate for each object in a flexible and open-ended way.

Beyond AR object inventory and natural language queries, several other applications make good use of the automatic semantic segmentation and indexing that our system provides. For example, users of our demo application enjoyed creating virtual copies of furniture in the scanned environment and probing where else in the room they could be placed in terms of fit, or for testing if the acquisition of additional matching furniture would prohibitively clutter the environment. 

\textbf{Design choices.} We have mostly tested our infrastructure via Magic-Leap-2-based optical-see-through AR, but we have also started experimenting with Meta Quest-3 and Quest Pro headsets. The system uses RGB/D data and camera poses for room reconstruction and only poses and pointers for interaction. Thus the user interface generalizes seamlessly to handheld devices such as iPads or video pass-through headsets. Video pass-through AR, with its more stringent occlusion capability, enables additional operations such as ``removal'' or ``modifications'' of physical objects. In particular, illusions similar to those described in \cite{kari_21_transformr, kari_23_scene_responsiveness} could be realized without the need to manually pre-mapping the environment. 

While working on our pipeline and prototype application, we experimented extensively with LLM APIs and local open-source LLMs. In this work, for the consideration of data privacy and efficiency, we decided against their use for the following reasons: a)~The most capable LLMs remain cloud-based, making them problematic for privacy-sensitive AR applications; b)~Although the throughput of LLM services is approaching real-time~\cite{llm_leaderboard}, the requirement of network traffic and added latency add friction to real-world AR use cases. Our CLIP-based solution provides context-agnostic features that greatly benefit computational efficiency and flexibility for downstream tasks – the expensive scene reconstruction happens once to support unlimited numbers of queries and object learning. The CLIP features sufficiently support the example applications while relying only on local computation. Most importantly, the user's data, like the scanned environment and the user preference model, all remain on the local storage.

\textbf{Limitations and future work.}\label{sec:limits} We implemented our proof-of-concept system with several deep-learning models that can run simultaneously on a single NVIDIA RTX 3090. The PC + Magic Leap 2 setup can easily be configured to a modern version of a portable backpack system such as the ``touring machine''~\cite{feiner1997touring}. Despite the encouraging results, these design choices: 1) constrained the linguistic understanding demonstration to spatial search in response to text queries instead of ``actual conversations with physical spaces'', and 2) limited the system's 3D object segmentation and object inventory feature to the ~100 categories of common stuff and things defined in the COCO dataset \cite{lin_microsoft_2014, kirillov2019panoptic}. 

To achieve true open-vocabulary 3D object segmentation, not limited by pre-defined categories is straightforward, and we did implement it, but did not optimize it for real-time performance. One solution is to adopt SegmentAnything~\cite{kirillov2023segany} to identify object boundaries and then use LMMs such as LLaVA~\cite{liu2023improvedllava} or GPT-4V~\cite{openai2024gpt} for neural vision-language feature extraction and image description, labeling, or captioning. 

Our current multimodal 3D scene model fusion approach opens new possibilities for context-aware AR interactions, such as responding to a natural-language query with a heatmap in AR (see \Cref{fig:queries}). Looking into the future of pervasive AR, we believe the ability to have a ``back-and-forth conversation'' with a physical space would be an attractive application for spatial computing. Think about asking your AR/AI system where in your backyard the best spots are to hang a hammock, or imagine you are a property manager and your conversational logging agent will proactively point out areas where closer inspection is needed based on previous findings spotted automatically during your current walk-through. 

Conversing with LLMs in text (and even images) is now as easy as texting, thanks to open-source and commercial solutions~\cite{touvron_llama_2023-1, openai2024gpt, llama3}. However, unlike short conversations, conversing with physical spaces requires us to ``tokenize the 3D model'' and feed a large amount of ``tokens'' or feature representations into LMMs/LLMs to get a spatially meaningful response. This vision is theoretically possible with several tweaks based on our current implementation, yet there are several challenges worth considering:

$\bullet$ What is the best approach to ``tokenize a $3m \times 3m$ space'' that can maintain its spatial meaning and at the same time keep a good balance between accuracy and efficiency (fewer tokens)?

$\bullet$ Existing LLMs have a limited context window--even with Google Gemini's seemingly large 1 million tokens window size~\cite{gemini_window}, sending the entire 3D context to an LLM repeatedly is not feasible. 

$\bullet$ The throughput of LLM  services (the number of output tokens per second) is approaching real-time for short contexts~\cite{llm_leaderboard}. Yet, considering the number of tokens required for 3D contexts, it remains a challenging task for real-time AR applications.

Our current intelligent object inventory implementation was designed to only recognize objects that are missing or unchanged in the physical world, but identifying ``insertions'' or previously unseen new objects could be a useful feature for real-world use cases. We discussed earlier that this limitation comes about because the training strategy is optimized for high accuracy for objects that appeared in previous 3D models. While we have seen more false positives than true positives in new object detection in ongoing experiments, inspirations from out-of-distribution detection~\cite{yang2021oodsurvey} research could point us to potential solutions. Differentiating two or more identical objects poses a significant challenge for both humans and machines. By applying everyday human tricks, such as utilizing relative spatial relationships between objects, we may teach machines to enhance such capabilities. Our open-source implementation highlights areas that may benefit from future improved solutions and quantitative evaluations.

\textbf{Implications.} In this work, we demonstrate new context-aware AR interfaces and applications by proposing a multimodal 3D fusion pipeline that integrates neural vision-language features into existing geometric and semantic representations of physical spaces. Our prototype intelligent inventory system re-identifies physical objects in AR and holds the promise of enhanced personal space management, team collaboration and information exchange, and even asset management. The heightened temporal awareness of the physical spaces could help tackle the issue of ``change blindness'' \cite{simons1997change}. The ability to perform 3D spatial searches using natural language can enable more intuitive AR interfaces -- conversing with physical spaces for interior design suggestions, safety inspection visualizations, or personalized navigation.

\acknowledgments{Support for this work was provided by ONR grant N00014-23-1-2118 as well as NSF grants IIS-2211784 and IIS-1911230.}

\bibliographystyle{abbrv-doi-hyperref-narrow}

\bibliography{template}
\end{document}